\documentstyle[twocolumn,osa]{revtex} 
\input epsf.sty

\def\fullsquare{\vrule height 4pt depth -1pt width 3pt\kern 0.1em}

\begin{document}

\title{Frequency tuning of the whispering gallery modes of silica
microspheres for CQED and spectroscopy}
\author{%
 Wolf von Klitzing\thanks{Present address:\\
                        Dipartimento di Fisica, Universit\`a di Roma ``La Sapienza'',
                        P.le A. Moro 2 , 00185 Roma, Italy},
 Romain Long,
 Vladimir S. Ilchenko\thanks{Present address:\\
                    Jet Propulsion Laboratory, California Institute of Technology,
                    4800 Oak Grove Drive, Pasadena, California 91109-8099},
 Jean Hare, Val\'erie Lef\`evre--Seguin}

\address{Laboratoire Kastler Brossel, D\'epartement de Physique de l'Ecole Normale Sup\'erieure,\\
         24 Rue Lhomond, 75231 Paris Cedex 05, France}

\maketitle  

\begin{abstract}
We have tuned the whispering gallery modes of a fused silica
microresonator over nearly 1~nm at 800~nm, i. e. over 0.5~FSR or
$10^6$ linewidths of the resonator. This has been achieved by a
new method based on the stretching  of a two--stem microsphere.
The devices described below will permit new Cavity--QED
experiments with this high--Q optical resonator when it is
desirable to optimize its coupling to emitters with given
transition frequencies. The tuning capability demonstrated here is
compatible with both UHV and low temperature operation, which
should be useful for future experiments with laser cooled atoms or
single quantum dots.
\end{abstract}

\bigskip
\noindent     
Light in a dielectric microsphere can be confined to so called whispering
gallery modes (WGMs).  The mode volume can be as low as a few hundred cubic
wavelengths with quality factors above $10^{9}$ for fused
silica.\cite{BraginskyGorodetsky89,CollotLefevre93,GorodetskySavchenkov96}
 Due to these advantageous properties, glass microspheres have attracted much
interest in various fields ranging from unusual lasers
\cite{SandoghdarTreussart96,KlitzingJahier99} over CQED,
\cite{MabuchiKimble94,TreussartHare94} spectroscopy,\cite{NorrisGonokami97} and
non--linear electrodynamics.\cite{BraginskyGorodetsky89,TreussartIlchenko98} So
far the main default of the microspheres has been their fixed frequency.
Accidental coincidences between an atomic line and the fundamental transverse
whispering gallery mode are extremely rare both because the free spectral
range (FSR) of a microsphere is very large, of the order of 1~THz, and because
the resonance linewidth is very small, about 300~kHz.  Tuning the whispering
gallery mode spectrum whilst preserving their high Q's is therefore an
important experimental challenge.

\bigskip

A useful device should fulfill the following conditions. The tuning range
should be about the free spectral range (FSR) so that a mode with the desired
transverse distribution of light can be tuned into resonance with, e.g., an
atomic transition. The device should be exceedingly stable because a change of
only $10^{-7}$ of the desired tuning range would already shift the WGMs by one
line width. Good access to the sphere must be guaranteed in order to enable
coupling light in and out of the sphere and also to approach the sample to the
sphere it is to interact with (quantum dots, cold atoms, etc \dots ).
Furthermore the device should be readily producible and affordable. This is
especially important for potential applications, e.g. as gas detectors.
Ultra-high vacuum compatibility is also required if one wishes to couple
laser--cooled atoms to a microsphere\cite{TreussartHare94}. In some cases, it
maybe desirable to operate the tunable microsphere at liquid helium
temperatures.

There are two methods to tune the modes: temperature and strain
\cite{SchillerByer91,IlchenkoGorodetsky92,CollotLefevre93}. At first order,
both affect the mode resonance through the simple relation: $\Delta \nu/\nu =
-\Delta a/a -\Delta N/N$, where $a$ is the radius of the sphere and $N$ its
refractive index. The temperature dependence of the modes is about
$-2.5$~GHz/K so that it can only be used for fine tuning purposes. Strain
tuning on the other hand can cover a whole FSR if one achieves $\Delta \nu/\nu
\simeq 1/l$. For a 50~micrometers sphere, this means a deformation of about
0.2\% in the equatorial region and an axial strain of about 1\%, which is
still compatible with the elastic deformation tolerated by fused silica, as
demonstrated by experiments performed on silica fibers
\cite{HustonEversole93}. The first demonstration device used on a microsphere
was designed to compress the sphere with piezo-driven pliers
\cite{IlchenkoVolikov98}. About one quarter of the sphere protruded from the
device thus allowing coupling to the WGMs. For a sphere of a diameter of
160~$\mu$m, tuning over 150~GHz, nearly one half of an FSR, was achieved.
However, access to the sphere was very limited and the device was not usable
for spheres smaller than 100~$\mu$m.  This precludes experiments on bulk
samples with quantum dots (access), for instance, and on lasing in doped
silica microspheres (size).

Here a new method is presented in which the strain is applied to the sphere by
stretching it. Advances in the ability to manipulate silica glass with a
CO$_2$ laser allow us to produce spheres with two stems, one on each pole. The
strain on the microsphere can now be exerted by simply pulling on the ends of
the two stems which should not be too thin, so as to produce enough strain on
the sphere's equator. With a sphere having twice or three times the diameter
of the stems, the desired stretching of the sphere can still be reached within
the elastic limit of silica.

We have made two different devices. Both of them give access to a
half space around the sphere, the other half being occupied by
the coupling optics. Device \#1 is a modified version of the
squeezer mentioned above. The double-stemmed sphere is produced
as follows. As in our previous experiments, the starting material
is the core of an optical fiber or a thin fiber pulled in a flame
from a rod of pure synthetic silica. The silica fiber to be melted
and the CO$_2$ laser beam are both vertical.  The fiber is
maintained vertical with a 5--10~mg  weight attached to its lower
end. The laser beam (diameter $\approx$ 3~mm) is projected
upwards through a focusing lens. The short focal length of 25~mm
creates a small focal spot (diameter $\approx$60~$\mu$m )and a
strong vertical gradient in the intensity of the infrared
radiation. As a result, the fiber hanging down will only melt near
the focus. Due to surface tension forces, a microsphere will form
and remain attached to the non-heated parts of the wire upon
cooling down. Its ellipticity can be below 1\%. Shortly after its
fabrication, the double-stemmed microsphere is glued in between
the jaws of the tuning device. Spheres with a diameter down to
about 60~$\mu$m having a stem diameter of about 30~$\mu$m can be
used in this device.

\begin{figure}[h]
\centerline{\epsfclipon\epsfxsize=8cm\epsffile{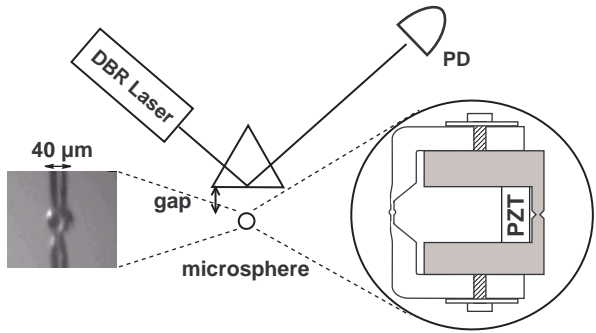}}
\caption{Simplified sketch of the experimental set-up. The WGM
modes of the microsphere are excited through a high index prism by
frustrated total internal reflection.  The resonances are detected
as dips on the laser light intensity coming out of the prism. The
right-side insert shows a side-view of the stretching device. The
left-side insert is a CCD image of a double-stemmed microsphere as
seen through our microscope. }
\end{figure}

The second device has been designed to work in a UHV environment on smaller
spheres. It consists of a U shaped base made from bronze which can be opened
and closed with a screw and a vacuum compatible low voltage PZT stack. As
shown in Fig.~1, two rods of pure silica bent in a flame are fixed with screws
on the jaws of the device so that they nearly meet at mid-height. Their tips
are then ground to the shape of pyramids with a tip to tip distance of
400~$\mu$m. Next the CO$_2$ laser is used to weld a short piece of silica
fiber across the gap. The sphere is then formed by melting the center of the
fiber with two counter-propagating CO$_2$ laser beams while the initial PZT
voltage is slowly relaxed. This production method has been used in an open
set-up but the last step could be performed directly in a high vacuum chamber
when necessary. Whatever the fabrication technique, the fusion process is
always observed in a stereomicroscope with a video camera and the alignment of
the CO$_2$ laser is carefully controlled by mounting the device and the laser
focusing lenses each on 3D translations stages. Care is taken to eliminate
residual tensions in the silica glass by gentle heating in the CO$_2$ laser
before and after the fusion process. The power of the CO$_2$ laser is
precisely controlled to avoid any excess heating which would cause unwanted
recrystallization of the silica and/or sublimation. This allows us to produce
a microsphere with two stems as short and thick as possible. This ensures
maximum strain and stability of the sphere (both mechanical and thermal). A
typical 40~$\mu$m microsphere is shown in the left-side insert of Fig.~1.

\begin{figure}[h]
\centerline{\epsfclipon\epsfxsize=8cm\epsffile{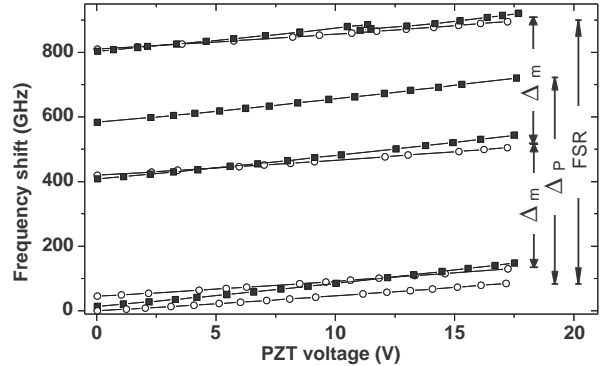}}
\caption{Frequency shift versus PZT voltage for several TM
(\fullsquare) or TE ($\circ$) whispering gallery modes Frequency
intervals, characterizing the WGM spectrum of the microsphere are
also shown, namely  the free-spectral range FSR, the separation
$\Delta_P$ between modes differing only in their polarizations,
and $\Delta_m$ the frequency difference due to a change by one
unit of $m$. TM modes are, as expected, more sensitive to stress
than TE modes by a factor 1.6. }
\end{figure}

The tuning of the whispering gallery modes was studied with a set-up described
previously.\cite{CollotLefevre93}
 Light from a diode laser is coupled into the sphere through a SF11-glass
prism. We studied first device \#1 using a 800~nm diode laser
stabilized with an external grating which had a maximum
continuous tuning range of 30~GHz with a linewidth of about
300~kHz. Like in Ref.~\onlinecite{IlchenkoVolikov98}, the average
tuning range of the modes in the sphere has been assessed by
continuously increasing the voltage at the PZT and observing the
modes passing through the frequency window scanned by the diode
laser. Assuming that there is no strong non-linearity in the
strain with respect to the PZT voltage the maximum tuning range
of this device was found to be 150~GHz, i.e about half of an FSR
for the 200~$\mu$m sphere under study. The stability of the WGM
frequency at a fixed PZT voltage was excellent: over a number of
days, the modes drifted by less than 10~GHz. On a short time
scale, the WGM frequency fluctutations were fully correlated to
the measured temperature variations. As usual in air, the modes
kept their initial Q value of about $10^8$ for several days.

The second device was studied with a narrow linewidth tunable DBR diode laser
(Yokogawa YL78XNL/S). The laser mode wavelength of these diodes can be tuned
continuously over 1~nm by simultaneously changing the diode and the grating
currents. This allowed us to follow the same WGM mode over 400~GHz. Fig.~2
shows the frequency shift of several WGM modes versus the PZT voltage for a
80~$\mu$m sphere. Here the voltage range was limited on purpose to stay within
the elastic limit and hence well below the maximum tuning range. We checked on
the  observed shift that the deformation was perfectly reversible upon
decreasing  the PZT voltage. The stability of the deformation was excellent,
as well as the repeatability of these results. The well-known
quasi-periodicity of the WGM mode spectrum is exhibited by marking three
characteristic frequency differences: FSR = 810~GHz is the free spectral range
of the sphere (diameter 80~$\mu$m), $\Delta_P =580$~GHz is the expected
interval between TE and TM modes with the same quantum numbers, $n$, $l$, $m$,
and finally $\Delta_m = 375$~GHz is the interval between modes which only
differ by one unit in $m$, the azimuthal order number.\cite{LaiLeung90-1} It
corresponds to an effective ellipticity of about 50\%, in agreement with a CCD
image of the sphere. The slope of the lines is 5~GHz/V for TE modes and
8~GHz/V for TM modes. The TM/TE slopes ratio is 1.6. It is close to 1.75, the
ratio expected for a perfect cylinder \cite{HustonEversole93}. This is
consistent with the large ellipticity of this resonator. The axial strain
derived from these measurements is $\epsilon \simeq 6\ 10^{-5}$ per Volt. This
yields an overall deformation of the sphere  with its two stems of 0.25~$\mu$m
for 10~V applied on the PZT, to be compared to the 0.5~$\mu$m/10~V
specification of the stack.

Fig.~3 shows the maximum excursion reached on this 80~$\mu$m sphere for a TE
and TM whispering gallery mode. When the PZT voltage was further increased up
to 42~V, the onset of plastic deformation was observed as a slip of the
frequency of the modes shortly before one of the stems severed.

\begin{figure}[h]
\begin{center}
\centerline{\epsfclipon\epsfxsize=8cm\epsffile{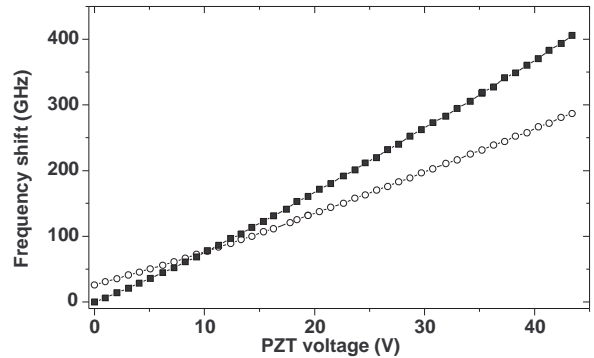}}
\caption{Frequency shift for  a pair of TM (\fullsquare) and TE
($\circ$) modes followed continuously over the maximum applied
stress. }
\end{center}
\end{figure}

To summarize, we have developed two devices which both have allowed us to tune
whispering gallery modes over half a FSR. In both cases the maximum quality
factor of Q=$10^9$ can be reached reproducibly and is preserved for a number
of days under standard laboratory conditions. Device \# 1 is easier to make
and it is more appropriate to work with moderate size spheres (60 to 500
micrometers) while device \# 2 is compatible with smaller spheres although it
needs more skill to get a small ellipticity. Now that microspheres can be
considered as tunable resonators, we believe that the devices presented here
open the way to several applications like single mode microlasers, tunable
filters, gas detectors and other cavity QED experiments with adsorbed
molecules, cold atoms grazing the sphere or quantum dots in semiconductor
samples.

{\bf Acknowledgements} ~This work has benefited from the financial
support of a  CEE TMR network contract \# ERBFMRXCT960066
"Microlasers and Cavity QED".


\begin{references}
\bibitem{note}
The precise design of the device should give a lever effect of about 5 at the
jaws of the device. It helps to compensate the imperfection of the device in
the region where the final welding of the upper stem was made.
\end{references}


\end{document}